\documentclass[10pt,conference]{IEEEtran}
\IEEEoverridecommandlockouts
\usepackage{balance}
\usepackage{amssymb}
\usepackage{stmaryrd}
\usepackage{cite}
\usepackage{color}
\usepackage{multirow}
\usepackage{graphicx,times}
\usepackage{epstopdf}
\usepackage{indentfirst}
\usepackage{CJK}
\usepackage{amsmath}
\usepackage{amsfonts}
\usepackage{txfonts}
\usepackage{mathrsfs}
\usepackage{theorem}
\usepackage{float}
\usepackage{subfigure}

\newtheorem{proposition}{Proposition}

\begin{document}
\title{Energy Beamforming for Wireless Information and Power Transfer in Backscatter Multiuser Networks}

\author{\IEEEauthorblockN{Wenyuan Ma, Wei Wang{${^\ast}$}, Tao Jiang}\IEEEauthorblockA{School of Electronic Information and Communications, Huazhong University of Science and Technology}Email: \{wenyuan\_ma, weiwangw, taojiang\}@hust.edu.cn
\thanks{${^\ast}$The Corresponding author is Wei Wang (weiwangw@hust.edu.cn).}}

\maketitle

\begin{abstract}
Wirelessly powered backscatter communication (WPBC) has been identified as a promising technology for low-power communication systems, which can reap the benefits of energy beamforming to improve energy transfer efficiency. Existing studies on energy beamforming fail to simultaneously take energy supply and information transfer in WPBC into account. This paper takes the first step to fill this gap, by considering the trade-off between the energy harvesting rate and achievable rate using estimated backscatter channel state information (BS-CSI). To ensure reliable communication and user fairness, we formulate the energy beamforming design as a max-min optimization problem by maximizing the minimum achievable rate for all tags subject to the energy constraint. We derive the closed-form expression of the energy harvesting rate, as well as the lower bound of the ergodic achievable rate. Our numerical results indicate that our scheme can significantly outperform state-of-the-art energy beamforming schemes. Additionally, the proposed scheme achieves performance comparable to that obtained via beamforming with perfect CSI.	        
\end{abstract}

\maketitle
%\vspace{-0.5cm}
\section{Introduction}
Wirelessly powered backscatter communication (WPBC), a kind of wireless transmissions by modulating and reflecting ambient wireless signals, emerges as a promising technology to achieve ultra-low-power communication for a broad range of the Internet of Things (IoT) applications, such as smart city, connected health, and smart farming \cite{han2016wirelessly}. Although achieving transmissions with power consumption of $\mu$Ws via ambient wireless signals from television and WiFi communications, WPBC is limited to short ranges due to the small amount of energy harvested by a backscatter tag and low signal power which further attenuates significantly after reflection.

By forming sharp energy beams towards target users, energy beamforming is an attractive solution to improve communication ranges. Growing attempts have been devoted to exploring the merits of energy beamforming in WPBC \cite{long2017transmit,yang2015multi}. Despite the fact that much has been understood through these studies, their beamforming designs fail to take the fundamental trade-off between energy supply and information transfer into account. These studies either only focus on optimizing the transmit beamforming to maximize the sum rate of a cooperative WPBC system\cite{long2017transmit}, or design energy beamforming just to maximize a total utility of the harvested energy \cite{yang2015multi}. Nevertheless, there are different optimal beams
for the achievable rate and energy harvesting rate. The reason lies in the fact that the energy harvesting rate at the tag only depends on the forward channel (i.e., transmitter-to-tag), while the achievable rate at the reader depends on the backscatter channel (i.e., transmitter-to-tag-to-receiver). As a consequence, energy beamforming that only maximizes harvested energy cannot guarantee the achievable rate to retain reliable communications. Likewise, energy beamforming that only maximizes data rate cannot ensure the energy supply.

 \begin{figure}
 	\centering
 	\includegraphics[width=0.95\linewidth]{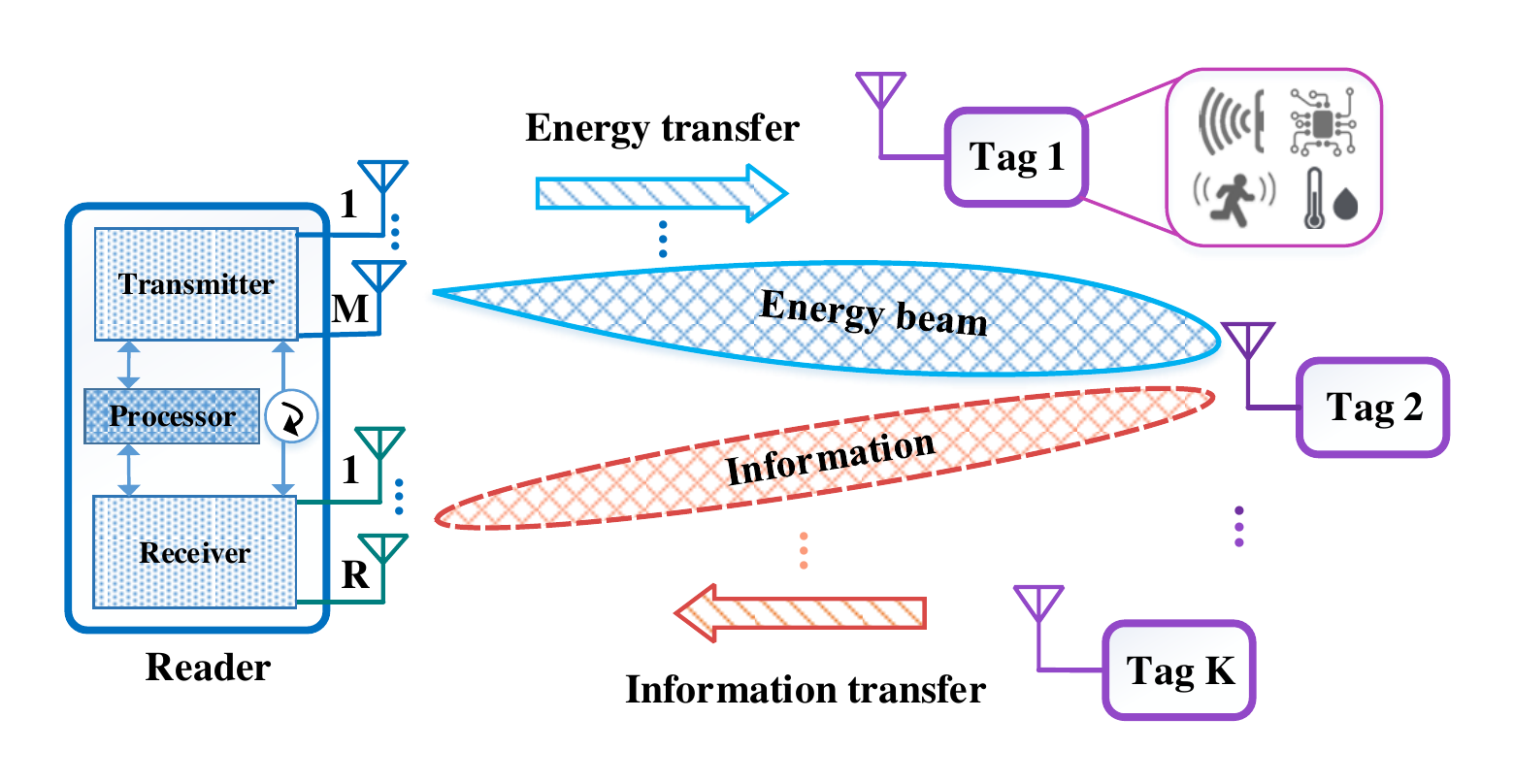}
 	\caption{A WPBC network with energy beamforming.}
 	\label{fig:figure1}
 \end{figure}

To overcome the above limitation, we take the first step to investigate the energy beamforming design that considers the trade-off between energy supply and information transfer via the estimated backscatter channel state information (BS-CSI). As shown in Fig.~\ref{fig:figure1}, this paper considers a general WPBC network where a reader with massive antennas transmits energy to multiple single-antenna tags, and tags transmit information to the reader by backscatter modulation. To ensure reliable communication and user fairness, we formulate the beamforming design as a max-min optimization problem by maximizing the minimum achievable rate for all tags subject to tags' power consumption constraints. To solve this problem, we analyze the energy harvesting rate and achievable rate via beamforming using the estimated BS-CSI.  

The major contributions are summarized as follows.
\begin{itemize}
	\item 	We propose a beamforming scheme for energy supply and information transfer in WPBC using the estimated BS-CSI. To the best of our knowledge, this is the first work that considers the trade-off between the energy harvesting rate and the achievable data rate in WPBC.
\end{itemize}
\begin{itemize}
	\item 	We obtain the analytical expressions of the energy harvesting rate and lower bound on the ergodic achievable rate with the estimated BS-CSI, considering the effect of unknown forward and backward channels.
\end{itemize}
\begin{itemize}
	\item 	Our numerical results indicate that the minimum achievable rate of the proposed scheme is significantly higher than that of previous energy beamforming schemes, while slightly lower than that obtained by energy beamforming with perfect CSI. 
\end{itemize}

The rest of this paper is organized as follows. Section \uppercase\expandafter{\romannumeral 2} establishes the system model. Section \uppercase\expandafter{\romannumeral 3} elaborates on the energy beamforming design for energy supply and information transfer. Section \uppercase\expandafter{\romannumeral 4} analyzes the harvested energy and achievable rate of the proposed scheme. Next, Section \uppercase\expandafter{\romannumeral 5} shows the numerical results. Section \uppercase\expandafter{\romannumeral 6} presents the state of the art. Finally, Section \uppercase\expandafter{\romannumeral 7} concludes our paper.

\section{System Model}
As shown in Fig.~\ref{fig:figure1}, we consider wireless energy transfer (WET) from a monostatic reader with $M$ antennas to $K$ single-antenna tags and backscatter communication between tags and the reader with another $R$ antennas for a WPBC system. Each tag harvests energy for circuit consumption and transmits information to the reader by backscatter modulation. 

We assume transmission over time blocks. The duration of each time block is $T$ symbol periods, each of which is shorter than the coherent interval. Each time block consists of a channel estimation (CE) slot, which lasts $\alpha$ symbol periods, followed by a simultaneous wireless energy and information transfer slot. First, the reader sends pilots and leverages backscattered signals to estimate the BS-CSI associated with each tag. Subsequently, the reader performs energy beamforming to transmit wireless energy and provide carrier signals for all tags. Simultaneously, tags transfer information to the reader. 

\subsection{Channel Model}
Let $\textbf{h}_{k}^{f}=\left[h_{1k}^{f}, h_{2k}^{f}, ... ,h_{Mk}^{f} \right]^T\in\mathbb{C}^{M\times1}$, $\textbf{h}_{k}^{b}=\left[h_{k1}^{b}, h_{k2}^{b}, ... , h_{kR}^{b} \right]^T\\\in\mathbb{C}^{R\times1}$ be the forward and backward channel vector, respectively. $h_{mk}^{f}$ is the forward channel coefficient between the $m$-th transmit antenna of the reader and tag $k$, and $h_{kr}^{b}$ is the backward channel coefficient between the $k$-th tag and the $r$-th receive antenna of the reader. We consider flat Rayleigh-fading channel model. Thus, we assume that $h_{mk}^{f} $, $h_{kr}^{b}$ are independent and both distributed as $\mathcal{CN}\left( 0, \beta_{k }\right)$, where $\beta_{k}$ denotes the path loss of the channel between the reader and tag $k$. $\beta_{k}$ is constant over the coherent interval and known at the reader. Let $\textbf{H}_{k}=\textbf{h}_{k}^{b}\textbf{h}_{k}^{fT}$ denote the $R\times M$ backscatter channel matrix of tag $k$, where $[\textbf{H}_{k}]_{mr}=h_{mkr}=h_{mk}^{f}h_{kr}^{b}$ is the backscatter channel coefficient, which is a combination of the channel from the $m$-th transmit antenna to tag $k$ with the channel from tag $k$ to the $r$-th receive antenna of the reader.

\subsection{Backscatter Channel Estimation}
In the backscatter CE slot, the reader estimates the channel matrix of each tag by controlling tags to reflect the pilot signals in sequence\cite{yang2015multi}. When estimating the backscatter channel of the $k$-th tag, the reader transmits orthogonal pilot sequences $\textbf{G} \textbf{B}^{1/2}\in{\mathbb{C}^{D\times M}}$ by $M$ transmit antennas, where $\textbf{G}\in{\mathbb{C}^{D \times M}}$ satisfies $\textbf{G}^{H}\textbf{G}=\textbf{I}_{M} $, and $\textbf{B}\in{\mathbb{C}^{M \times M}}$ is a diagonal matrix with $Dp_{ce}$ as element. $p_{ce}$ and $D$ represent the power of each antenna transmitting a pilot sequence and the length of pilot sequence $(D\ge M)$, respectively. Then, $K$ tags reflect the pilot signals with a reflection coefficient $\delta_{k}\in{\mathbb{C}}$ in sequence. 

The reader can eliminate the direct-link influence by employing self-interference cancellation techniques or frequency shifting. Thus, the received signal at the reader is given by
\begin{equation}
\begin{aligned}
\textbf{Y}_{k}^{CE} &=\sqrt{\delta_{k}}\textbf{H}_{k}(\textbf{G} \textbf{B}^{1/2})^{T}+\textbf{N}_{k},
\end{aligned}
\end{equation} 
where $\textbf{N}_{k}$ is an $R\times D$ matrix with independent and identically distribute (i.i.d.) elements and $\left[ \textbf{N}_{k}\right]_{mr} = n_{k,mr} \sim \mathcal{CN}\left( 0, \sigma^2\right)$.

Given $\textbf{Y}_{k}^{CE}$, the least-square (LS) estimate of $\textbf{H}_{k}$ can be expressed as
\begin{equation}
\begin{aligned}
\widehat{\textbf{H}}_{k}^{LS}&=\textbf{Y}_{k}^{CE} \delta_{k}^{-1/2}\textbf{G}^{*}\textbf{B}^{-1/2}\\ &=\textbf{H}_{k}+\textbf{N}_{k}\delta_{k}^{-1/2}\textbf{G}^{*}\textbf{B}^{-1/2}.
\end{aligned}
\end{equation}  
Denote the estimation error by $\widetilde{\textbf{E}}_{k}\triangleq \widehat{\textbf{H}}_{k}^{LS}-\textbf{H}_{k} $, where $ \tilde{e}_{k,mr}$, $\hat{h}_{mkr}^{LS}$, and $h_{mkr}$ represent the $m$-th row , $r$-th column element of $\widetilde{\textbf{E}}_{k}$, $\widehat{\textbf{H}}_{k}^{LS}$, and $\textbf{H}_{k} $, respectively. $\widetilde{\textbf{E}}_{k}$ is independent of $\widehat{\textbf{H}}_{k}^{LS}$. Thus, we can obtain $\tilde{e}_{k,mr}\sim \mathcal{CN}\left( 0, \frac{\sigma^2}{Dp_{ce}\delta_{k}}\right)$.

However, energy beamforming needs to obtain forward channel state information (F-CSI). Therefore, we use the backward channel state information (B-CSI) as an unknown parameter to obtain the desired F-CSI through BS-CSI. Mathematically, we define random vector $\widehat{\textbf{h}}_{k}^{f}\triangleq \widehat{\textbf{h}}_{kr}/ h_{kr}^{b}$, where $ \widehat{\textbf{h}}_{k}^{f}=[\hat{h}_{1k}^{f}, \hat{h}_{2k}^{f}, ..., \hat{h}_{Mk}^{f}]^T$, $\widehat{\textbf{h}}_{kr}=[\hat{h}_{1kr}, \hat{h}_{2kr}, ... ,\hat{h}_{Mkr}]^T$, $r\in\{1, 2, ..., R\}$, and the error  $\textbf{e}_{k}\triangleq \widehat{\textbf{h}}_{k}^{f}-\textbf{h}_{k}^{f}$. 
Then, we can obtain the element of error that follows complex Gaussian distribution with zero mean and variance
\begin{equation}\label{6}
\sigma_{e,kr}^2(h_{kr}^{b})=\frac{\sigma^2}{|h_{kr}^{b}|^2Dp_{ce}\delta_{k}}                                                             =\frac{K\sigma^2}{|h_{kr}^{b}|^2\alpha p_{ce}\delta_{k}}.
\end{equation}

Based on the above analysis, we can obtain the distributions with $h_{kr}^{b}$ as a condition for subsequent analysis as follows:
\begin{equation}\label{8}
\begin{aligned}
\widehat{\textbf{h}}_{kr}|h_{kr}^{b} &\sim \mathcal{CN}\left(\textbf{0}_{M},|h_{kr}^{b}|^2\left( \beta_{k}+\sigma_{e,kr}^2(h_{kr}^{b})\right) \textbf{I}_{M}\right) \\
\widehat{\textbf{h}}_{k}^{f}|h_{kr}^{b} &\sim \mathcal{CN}\left(\textbf{0}_{M},\left( \beta_{k}+\sigma_{e,kr}^2(h_{kr}^{b})\right) \textbf{I}_{M}\right) \\
\textbf{h}_{k}^{f}|\widehat{\textbf{h}}_{kr},h_{kr}^{b} &\sim \mathcal{CN}\left(\frac{\beta_{k}\widehat{\textbf{h}}_{k}^{f}}{\beta_{k}+\sigma_{e,kr}^2(h_{kr}^{b})}, \frac{\beta_{k}\sigma_{e,kr}^2(h_{kr}^{b})}{\beta_{k}+\sigma_{e,kr}^2(h_{kr}^{b})}\textbf{I}_{M} \right). 
\end{aligned}
\end{equation} 

\section{Energy Beamforming for Wireless Energy and Information Transfer}
\subsection{Energy Transfer}
The reader transmits a carrier signal to $K$ tags concurrently. To improve the energy transmission efficiency, we use the estimated BS-CSI for beamforming. The reader transmits a signal $u$ with $\mathbb{E} \left\lbrace \left| u \right|^2  \right\rbrace=p$, and the average transmit power is $w$. Thus, we can obtain $p=\frac{wT -\alpha p_{ce}}{T-\alpha} $. The reader uses a weighted sum of the normalized estimated BS-CSI, since it is proved to be asymptotically optimal for WET\cite{zhang2013mimo}. The beamformer is denoted as  $\boldsymbol{\Phi}=\sum_{k=1}^{K}\sqrt{\zeta_{k}}\frac{\hat{\textbf{h}}_{kr}^*}{\|\hat{\textbf{h}}_{kr}\|}$, where $\zeta_{k}\in{[0,1]}$ $\forall \, k$ such that $\sum_{k=1}^{K}\zeta_{k}=1$. So the signal emitted by the reader is given as $u\boldsymbol{\Phi}$, and the received signal at tag $k$ is given by
\begin{equation}\label{12}
b_{k} =u\boldsymbol{\Phi}^T\textbf{h}_{k}^{f} + n_{k},
\end{equation}
where $n_{k}$ denotes the noise and  $n_{k}\sim \mathcal{CN}\left( 0, \sigma_{k}^2\right)$.

We assume that the noise energy cannot be harvested \cite{yang2015multi}. Then the incident signal power of tag $k$ can be represented as 
\begin{equation}\label{sec2:incident signal power}
P_{Ik}=\mathbb{E} \left\lbrace \left| u \right|^2 |\boldsymbol{\Phi}^T\textbf{h}_{k}^{f}|^2  \right\rbrace
=p\mathbb E\left\lbrace|\boldsymbol{\Phi}^T\textbf{h}_{k}^{f}|^2 \right\rbrace.
\end{equation}
When the tag is activated to reflect the incoming signal, The rectifier converts the $(1-\delta_{k})$ fraction of incident radio frequency signal into a direct current (DC)\cite{lu2018wireless}. Otherwise, the whole incident signal power is rectified to DC.

To simplify the analysis, we assume that the harvested energy scales linearly with the input power \cite{khan2018optimization}. Thus, the instantaneous energy harvesting rate when the tag is activated is given by 
\begin{equation}\label{energy harvesting rate}
P_{Ek}=\eta(1-\delta_{k})P_{Ik},
\end{equation}
where $\eta\in\left(0,1\right]$ is the rectifier efficiency. 

\subsection{Information Transfer}
When the instantaneous energy harvesting rate exceeds the circuit power consumption rate, tags reflect and modulate the incident carrier signal. Then the received signal vector from tag $k$ can be written as
\begin{equation}\label{16}
\begin{aligned}
\textbf{y}_{k}& =\sqrt{\delta_{k}}\textbf{h}_{k}^{b}b_{k}s_{k}+\textbf{z}_{k}\\   &=\sqrt{\delta_{k}}us_{k}\textbf{h}_{k}^{b}\textbf{h}_{k}^{fT}\boldsymbol{\Phi} + \textbf{v}_{k},
\end{aligned} 
\end{equation}
where $\delta_{k}$ and $s_{k}$ are the reflection coefficient and the backscatter signal of tag $k$, respectively. We assume $s_{k}$ is distributed as $\mathcal{CN}\left( 0, 1 \right) $, and the noise $\textbf{v}_{k}$ is distributed as $\mathcal{CN}\left( \textbf{0}, \boldsymbol{\Sigma} \right) $ \cite{long2017transmit}.

So the received signal vector at the reader from $K$ tags can be represented as
\begin{equation}\label{17}
\vspace{-0.1cm}
\begin{aligned}
\textbf{y} = \sum_{k=1}^{K} \textbf{y}_{k} &=\textbf{H}^{b}\left[\left(\textbf{H}^{fT}u\boldsymbol{\Phi}\right)\circ\left(\boldsymbol{\Lambda}\textbf{s}\right)\right]+\boldsymbol{\mu} \\&=\textbf{H}^{b}\textbf{x}+\boldsymbol{\mu},
\end{aligned} 
\end{equation}
where $\circ$ represents Hadamard product, $\textbf{H}^{b}=[\textbf{h}_{1}^{b}, \textbf{h}_{2}^{b}, ... , \textbf{h}_{K}^{b}]$, $\textbf{H}^{f}=[\textbf{h}_{1}^{f}, \textbf{h}_{2}^{f}, ... , \textbf{h}_{K}^{f}]$, $\boldsymbol{\Lambda}=diag\{\sqrt{\delta_{1}}, \sqrt{\delta_{2}}, ... , \sqrt{\delta_{K}}\}$, $\textbf{s}=[s_{1}, s_{2}, ... , s_{K}]^T$, $\textbf{x}=[\sqrt{p_{1}}s_{1}, \sqrt{p_{2}}s_{2}, ... , \sqrt{p_{K}}s_{K}]^T$, the $p_{k} = \delta_{k} P_{Ik}$ denotes the reflect power, and $\boldsymbol{\mu}$ is a vector of additive white Gaussian noise, $\boldsymbol{\mu}$ has i.i.d. $\mathcal{CN}\left( 0, \tilde{\sigma}^2 \right)$ elements.

%So the received signal vector at the reader from $K$ tags can be represented as
%\begin{equation}\label{17}
%\begin{aligned}
%\textbf{y} =\sum_{k=1}^{K}\textbf{y}_{k}=\textbf{H}^{b}\textbf{x}+\boldsymbol{\mu},
%\end{aligned} 
%\end{equation}
%where $\textbf{H}^{b}=[\textbf{h}_{1}^{b}, \textbf{h}_{2}^{b}, ... , \textbf{h}_{K}^{b}]$, $\textbf{x}=[\sqrt{p_{1}}s_{1}, \sqrt{p_{2}}s_{2}, ... , \sqrt{p_{K}}s_{K}]^T$,  the $p_{k} = \delta_{k} P_{Ik}$ denotes as the reflect power, and the $\boldsymbol{\mu}$ is a vector of additive white Gaussian noise. The element of noise vector is distributed as $\mathcal{CN}\left( 0, \tilde{\sigma}^2 \right) $.

However, information detection needs to obtain the B-CSI. Therefore, we use F-CSI as an unknown parameter to obtain the desired B-CSI through BS-CSI. Mathematically, we define random vector $\widehat{\textbf{h}}_{k}^{b}\triangleq \widehat{\textbf{h}}_{mk}/ h_{mk}^{f}$, where $\widehat{\textbf{h}}_{k}^{b}=[\hat{h}_{k1}^{b}, \hat{h}_{k2}^{b}, ..., \hat{h}_{kR}^{b}]^T$,  $\widehat{\textbf{h}}_{mk}=[\hat{h}_{mk1}, \hat{h}_{mk2}, ... ,\hat{h}_{mkR}]^T$, $m\in\{1, 2, ..., M\}$, and the error  $\boldsymbol{\varepsilon}\triangleq \widehat{\textbf{H}}^{b}-\textbf{H}^{b}$. Furthermore, we can obtain the distributions of $
\widehat{\textbf{h}}_{mk}|h_{mk}^{f}$, $\widehat{\textbf{h}}_{k}^{b}|h_{mk}^{f}$, $\textbf{h}_{k}^{b}|\widehat{\textbf{h}}_{mk},h_{mk}^{f}$ according to the analytical results in Section \uppercase\expandafter{\romannumeral 2}-B.

The received vector after using a linear detector $\textbf{Q}$ to detect the information from all tags is given by
\begin{equation}
\begin{aligned}
\textbf{r} &= \textbf{Q}^{H} \textbf{H}^{b} \textbf{x} + \textbf{Q}^{H} \boldsymbol{\mu} = \textbf{Q}^{H} \left( \widehat{\textbf{H}}^{b} \textbf{x} - \boldsymbol{\varepsilon} \textbf{x} + \boldsymbol{\mu}\right). 
\end{aligned} 
\end{equation}
Let $r_{k}$ and $x_{k}$ be the $k$-th elements of $K \times 1$ vectors $\textbf{r}$ and $\textbf{x}$, respectively. $\textbf{q}_{k}$, $\widehat{\textbf{h}}_{k}^{b}$, $\boldsymbol{\varepsilon}_{k}$ be the $k$-th column of detector $\textbf{Q}$, $\widehat{\textbf{H}}^{b}$, $\boldsymbol{\varepsilon}$. Therefor, after using the linear detector, the received signal associated with the $k$-th tag can be written as 
\begin{equation}\label{key}
r_{k} = \textbf{q}_{k}^{H} \widehat{\textbf{h}}_{k}^{b} x_{k} + \sum_{i=1,i\neq k}^{K} \textbf{q}_{k}^{H} \widehat{\textbf{h}}_{i}^{b} x_{i} - \sum_{i=1}^{K} \textbf{q}_{k}^{H} \boldsymbol{\varepsilon}_{i} x_{i}+ \textbf{q}_{k}^{H} \boldsymbol{\mu}.
\end{equation}

The ergodic achievable rate of the information transmission from tag $k$ is given by
\begin{equation}\label{rate}
R_{k} \triangleq \mathbb{E} \left\lbrace \log_{2}\left( 1 +  \gamma_{k} \right) \right\rbrace,  
\end{equation}
where the signal-to-interference-plus-noise-ratio(SINR)
\begin{equation}\label{key}
\vspace{-0.2cm}
\begin{aligned}
\gamma_{k} & =   \frac{ p_{k} |\textbf{q}_{k}^{H} \widehat{\textbf{h}}_{k}^{b}|^2 }{\sum_{i=1,i\neq k}^{K}  p_{i} |\textbf{q}_{k}^{H} \widehat{\textbf{h}}_{i}^{b}|^2  + |\textbf{q}_{k}^{H} \textbf{q}_{k}|\sum_{i=1}^{K}  p_{i}  \sigma_{\epsilon,mi}^2(h_{mi}^{f})+ |\textbf{q}_{k}^{H} \textbf{q}_{k}|\tilde{\sigma}^2}.
\end{aligned} 
\end{equation}  

\subsection{Energy Beamforming Design}
We realize energy beamforming and information detection for energy supply and information transfer in WPBC using the estimated BS-CSI. We note that the energy harvesting rate at the tag relies on the forward channel while the achievable rate at the reader relies on the backscatter channel. Therefore, there are different optimal energy beams for achievable rate and harvested energy due to the difference between the forward and backscatter channels. We formulate the beamforming design as a max-min optimization problem to maximize the minimum rate for all tags by jointly optimizing the energy allocation weights $\boldsymbol{\zeta}=\left[\zeta_{1}, \zeta_{2},..., \zeta_{K} \right]$ of the beamformer $\boldsymbol{\Phi}$, the CE time $\alpha$ and transmit power of pilot $p_{ce}$, subject to the power consumption constraint as follows.
\begin{equation}\label{problem formulation}
\begin{aligned}
\qquad &\max\limits_{\boldsymbol{\zeta},\alpha, p_{ce} }\quad \min \limits_{1 \le k \le K} R_{k}\\
& \begin{array}{r@{\quad}r@{}l@{\quad}l}
s.t.&&\quad P_{Ek} \geq \rho, \quad \forall k \\
&&\quad \sum_{k=1}^{K}\zeta_{k}=1,\\
&&\quad 0 \leq \alpha \leq T,\\
&&\quad \zeta_{k}  \geq 0, \quad \forall k,
\end{array} 
\end{aligned} 
\end{equation}
where $\rho$ denotes the circuit power consumption rate in backscattering. This problem maximizes the throughput with user-fairness assurance. However, a closed-form analytic solution to \eqref{problem formulation} does not appear easy to find. To verify the performance improvement of the proposed scheme, we obtain the optimal resource allocation through numerical solver.

\newcounter{TempEqCnt}                         % 创建临时变量TempEqCnt
\setcounter{TempEqCnt}{\value{equation}} % 将当前公式序号 赋给TempEqCnt
\setcounter{equation}{16}                           % 当前公式序号变为x，x等于长公式应有的序号减1.
\begin{figure*}[ht]
	\begin{equation}\label{lower boound}
	\begin{aligned}
	R_{k} \ge \tilde{R}_{k} &\triangleq \log_{2}(1 + \tilde{\gamma}_{k})
	= \log_{2}\left(1 + \left(\mathbb{E} \left \lbrace \frac{\sum_{i=1,i\neq k}^{K}  p_{i} |\textbf{q}_{k}^{H} \widehat{\textbf{h}}_{i}^{b}|^2  + |\textbf{q}_{k}^{H} \textbf{q}_{k}|\sum_{i=1}^{K}  p_{i}  \sigma_{\epsilon,mi}^2(h_{mi}^{f})+ |\textbf{q}_{k}^{H} \textbf{q}_{k}|\tilde{\sigma}^2}  { p_{k} |\textbf{q}_{k}^{H} \widehat{\textbf{h}}_{k}^{b}|^2 }\right \rbrace \right)^{-1}\right).
	\end{aligned} 
	\end{equation} 
	\hrulefill
\end{figure*}
\setcounter{equation}{\value{TempEqCnt}} % 把TempEqCnt中存的公式序号赋回给当前公式序号
       
\section{Analysis on Energy Harvested Rate and Achievable Rate}
To optimize beams in a multiuser system as formulated in Eq. \eqref{problem formulation}, we derive analytical expressions of the energy harvesting rate and achievable rate by using energy beamforming with the estimated BS-CSI. Additionally, we analyze the effect of the unknown channels on the energy harvesting rate and achievable rate. For analytical tractability, we obtain the bounds on the energy harvesting rate and achievable rate.

\subsection{Energy Harvesting Rate}
Based on the energy beamforming, the energy harvesting rate \eqref{energy harvesting rate} can be derived as \cite{yang2015multi}
\begin{equation}\label{incident signal power}
\begin{aligned}
P_{Ek}=\eta(1-\delta_{k})p \beta_{k} \left[\zeta_{k}\left[ M-\left( M-1\right)\phi\left(h_{kr}^{b}\right) - 1\right] +1\right],\\
\end{aligned}
\end{equation}
where $\phi\left(h_{kr}^{b}\right)=\mathbb E_{h_{kr}^{b}}\left\lbrace\frac{1}{\frac{\beta_{k}|h_{kr}^{b}|^2\alpha p_{ce}  \delta_{k}}{K\sigma^2}+1}\right\rbrace$. 
The result of \eqref{incident signal power} is obtained by performing expectation over the unknown backward channel, which takes the fuzziness of $h_{kr}^{b}$ into account. To simplify the calculation, we derive the upper and lower bounds on the incident signal power. 

\begin{proposition}
	The boundary of the incident signal power of tag $k$, when energy bamforming is performed using the BS-CSI, is given by
	\begin{equation}
	\begin{aligned}
	P_{Ik}^{L}  &\triangleq p\beta_{k} \left[\zeta_{k} (M-1)\mathcal{L}_{1}(\alpha, p_{ce}) +1\right] < P_{Ik}(p_{ce}, p, \alpha, \zeta_{k})\\
	& < p\beta_{k} \left[\zeta_{k} (M-1) \mathcal{L}_{2}(\alpha, p_{ce})+1\right] \triangleq P_{Ik}^{U},
	\end{aligned}
	\end{equation}
where\\
 \indent$\mathcal{L}_{1}(\alpha, p_{ce})=\left[1-\frac{K\sigma^2}{\beta_{k}^2 \alpha p_{ce} \delta_{k}} \ln(1 + \frac{ \beta_{k}^2 \alpha p_{ce} \delta_{k}}{ K\sigma^2 })\right]$,\\
\indent$\mathcal{L}_{2}(\alpha, p_{ce})=\left[1-\frac{K\sigma^2}{2 \beta_{k}^2 \alpha p_{ce} \delta_{k}} \ln(1 + \frac{ 2 \beta_{k}^2 \alpha p_{ce} \delta_{k}}{ K\sigma^2 })\right]$.	
\end{proposition}
\begin{IEEEproof}
  Please refer to Appendix~A.
\end{IEEEproof}

\subsection{Ergodic Achievable Rate}
The exact expression for the rate $R_{k}$ in \eqref{rate} is analytically intractable. Since $f(x)=\log(1+ \frac{1}{x})$ is a convex function, by utilizing Jensen's inequality, the lower bound on the ergodic achievable rate is given by \eqref{lower boound} shown at the top of this page. The result of \eqref{lower boound} is obtained by performing expectation over the unknown backward and forward channels, which takes the effect of unknown channels into account.

\begin{proposition}
 With LS channel estimate, MRC detection, i.e. $\textbf{q}_{k} = \frac{\widehat{\textbf{h}}_{mk}}{\|\widehat{\textbf{h}}_{mk}\|}$, and $R \ge 2$, we can obtain an achievable rate as follows
 \setcounter{equation}{17}
 \begin{equation}
 \begin{aligned}
 \tilde{R}^{MRC}_{k} &= \log_{2}(1 + \tilde{\gamma}^{MRC}_{k}),\\
 \end{aligned} 
 \end{equation} 
 where $p_{k} = \delta_{k} P_{Ik}$, and the SINR
 \begin{equation}\label{SINR}
 \begin{aligned}
 \vspace{-0.5cm}
 &\tilde{\gamma}^{MRC}_{k} = \frac{p_{k} (R-1)\beta_{k}}{\left( 1 - \frac{K\sigma^2}{\beta_{k}^2 \alpha p_{ce} \delta_{k}} \exp(\frac{K\sigma^2}{\beta_{k}^2 \alpha p_{ce} \delta_{k}}) \Gamma(0,\frac{K\sigma^2}{\beta_{k}^2 \alpha p_{ce} \delta_{k}})\right) }\\
 & \frac{1}{\left( \sum_{i=1,i\neq k}^{K}  p_{i} \left(\beta_{i} + \frac{K \sigma^2}{\alpha p_{ce} \delta_{i}} \Gamma(0, \frac{\tau}{\beta_{i}}) \right) + \sum_{i=1}^{K}  p_{i} \frac{K \sigma^2}{\alpha p_{ce} \delta_{i}}\Gamma(0, \frac{\tau}{\beta_{i}}) + \tilde{\sigma}^2 \right)}.
 \end{aligned} 
 \end{equation} 
\end{proposition}
\begin{IEEEproof}
	Please refer to Appendix~B.
\end{IEEEproof}

We also note that the SINR appears analytically intractable. For analytical tractability, we obtain the lower bound on the SINR as \eqref{lower SINR} shown at the top of next page. We take the lower bound $\tilde{R}^{MRC}_{kL}$ as the achievable rate, and $P_{Ek}^{L}$ as instantaneous energy harvesting rate in Eq. \eqref{problem formulation} to complete the beamforming design.

\begin{figure*}[ht]
\begin{equation}\label{lower SINR}
\tilde{\gamma}^{MRC}_{kL} = \frac{\delta_{k} P_{Ik}^{L} (R-1)\beta_{k}}{\left( 1 - \frac{K\sigma^2}{2 \beta_{k}^2 \alpha p_{ce} \delta_{k}} \ln(1 + \frac{ 2 \beta_{k}^2 \alpha p_{ce} \delta_{k}}{ K\sigma^2 })\right) \left( \sum_{i=1,i\neq k}^{K}  \delta_{i} P_{Ii}^{U} \left(\beta_{i} + \frac{K \sigma^2}{\alpha p_{ce} \delta_{i} \beta_{i}} e^{-\frac{\tau}{\beta_{i}}} \ln(1+\frac{\beta_{i}}{\tau})  \right) + \sum_{i=1}^{K} \delta_{i} P_{Ii}^{U} \frac{K \sigma^2}{\alpha p_{ce} \delta_{i} \beta_{i}} e^{-\frac{\tau}{\beta_{i}}} \ln(1+\frac{\beta_{i}}{\tau}) + \tilde{\sigma}^2 \right)}.
\end{equation}
\hrulefill
\end{figure*}

\section{Numerical Results}
In this section, we verify the performance improvement of the proposed scheme through numerical simulations. We consider two tags and assume the duration of each time block is $T = 200$ symbol periods. We set the noise power of the backward link as $\tilde{\sigma}^2 = - 60$ dBm and noise power of the whole backscatter link as $\sigma^2 = - 90$ dBm. We also use the long-term fading model $\beta_{k} = \frac{0.0086}{4\pi d_{k}^2}$ \cite{yang2015multi}, where the distance $d_{1} = 4$ m and $d_{2} =6$ m. The energy conversion coefficient is $\eta = 0.65$ and the reflection coefficient of two tags is $\delta_{1} = \delta_{2} = 0.3 + 0.4i$ \cite{gang2018Modulation}. We consider the circuit power consumption rates as $\rho = 8.9~\mu W$ \cite{lu2018wireless}. 
\begin{figure}
	\centering
	\includegraphics[width=0.85\linewidth]{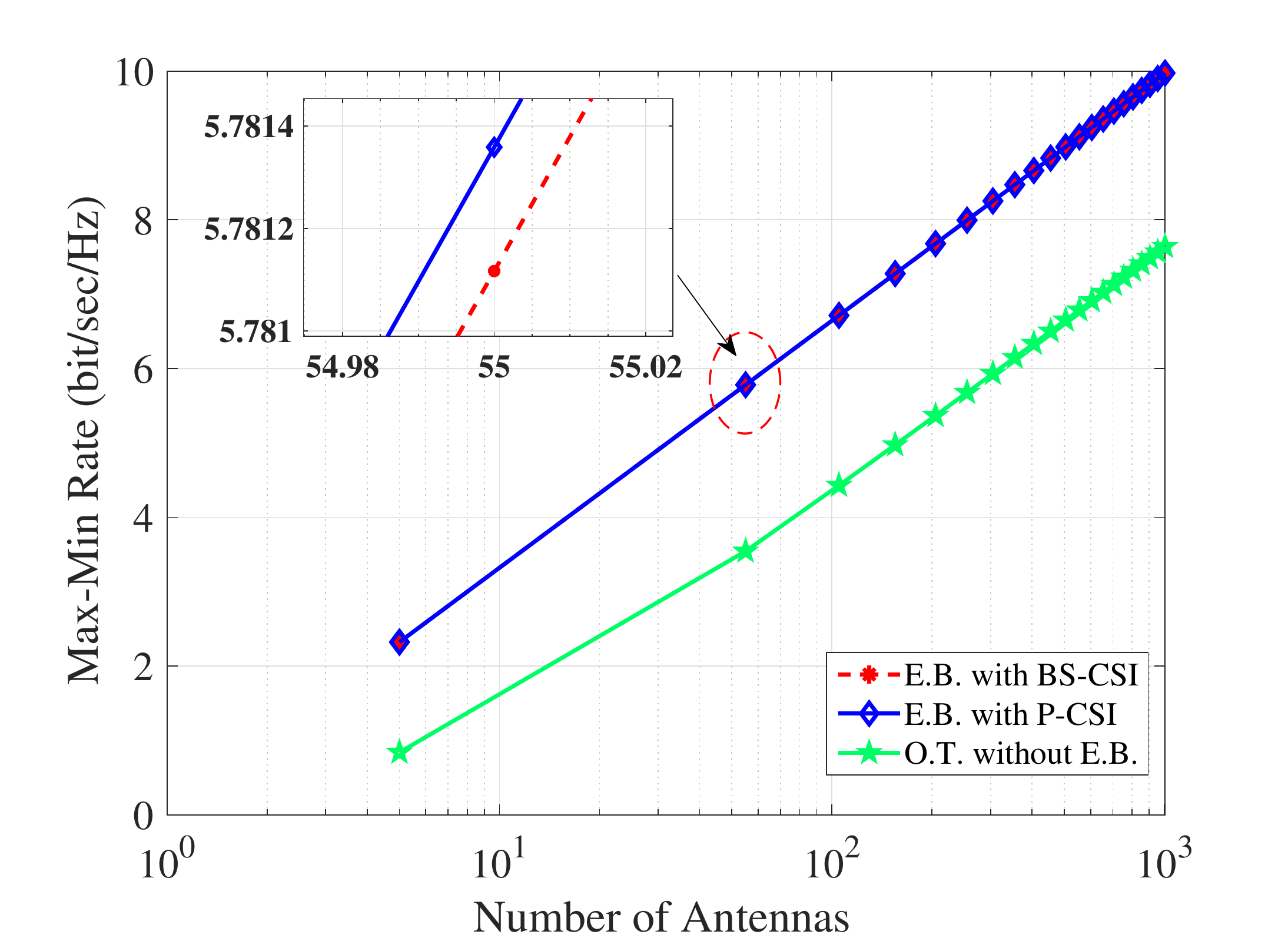}
	\caption{Maximal minimum achievable rate vs. $R$ for three schemes.}
	\label{fig:figure2}
\end{figure}
\begin{figure}
	\centering
	\includegraphics[width=0.85\linewidth]{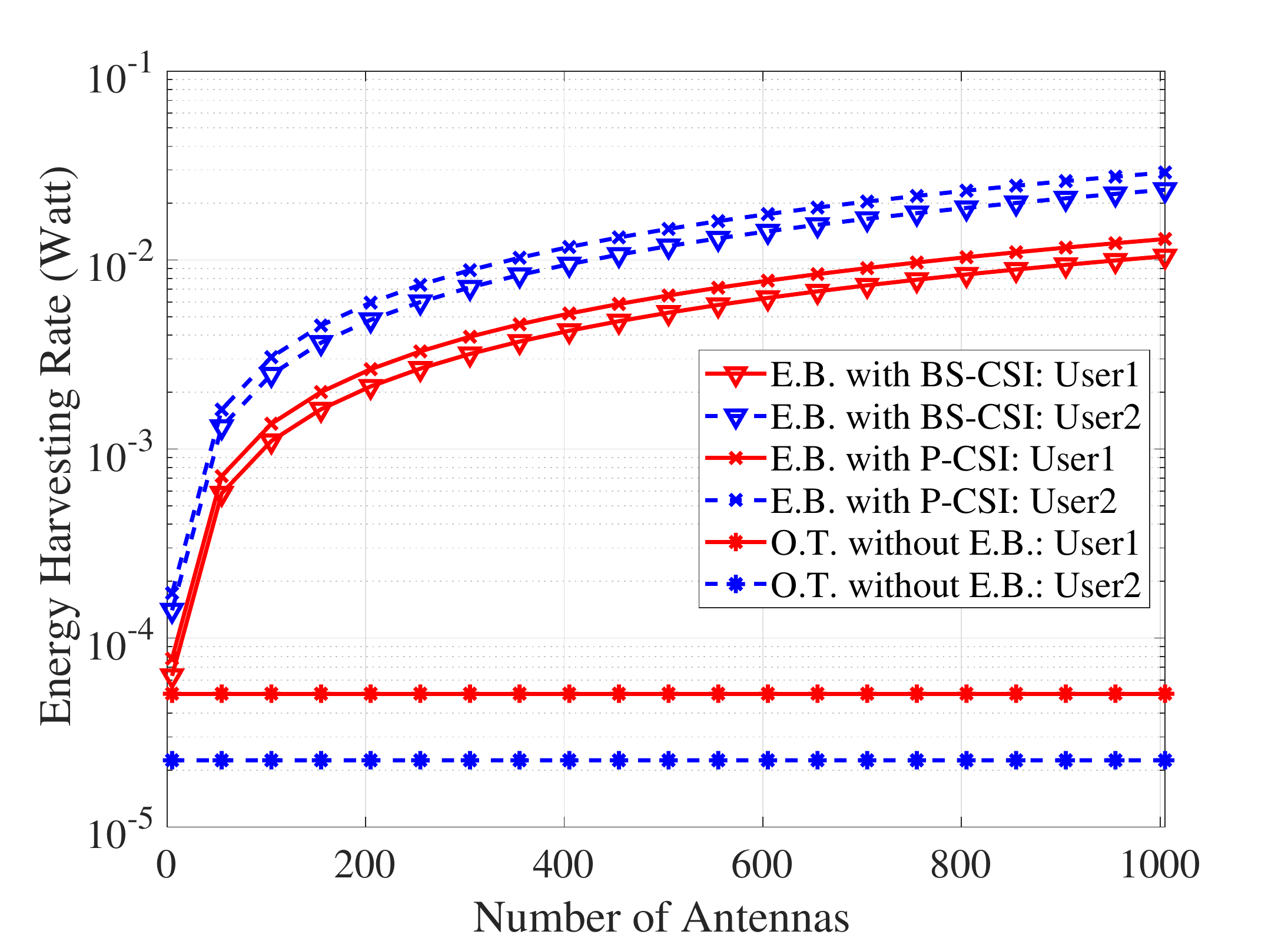}
	\caption{Energy harvesting rate vs. $M$ for three schemes.}
	\label{fig:figure3}
\end{figure}

To verify the performance gain of the proposed scheme, we give two benchmarks, i.e., the case of beamforming with perfect CSI and the case of omnidirectional transmission. In the first case, we assume the reader has perfect knowledge of F-CSI for energy beamforming and B-CSI for information detection. Thus, the step of the CE can be removed. In the second case, the reader performs omnidirectional transmission without energy beamforming. Channel estimation is only used for information detection. 

Figs.~\ref{fig:figure2} and \ref{fig:figure3} show the efficiency of the proposed scheme. We first compare the achievable rate to the two benchmarks. As shown in Fig.~\ref{fig:figure2}, we can observe that the max-min rate of the proposed scheme is significantly higher than that of omnidirectional transmission, while slightly lower than that obtained by energy beamforming with perfect CSI. From Fig.~\ref{fig:figure3}, we can also observe that the energy harvesting rate of the proposed scheme has a large gain compared to that of omnidirectional transmission, only a slight loss compared to that obtained using perfect CSI. Furthermore, as $M$ increases, the energy harvesting rate by using beamforming increases, while the energy harvesting rate by omnidirectional transmission remains as a small constant. Due to the extremely low energy harvesting rate of omnidirectional transmission, it is difficult to activate the tag for communication when the transmission power at the reader is low. However, beamforming can greatly increase the energy harvesting rate to activate tags and ensure uninterrupted backscattering of tags. 

\begin{figure}
	\centering
	\includegraphics[width=0.85\linewidth]{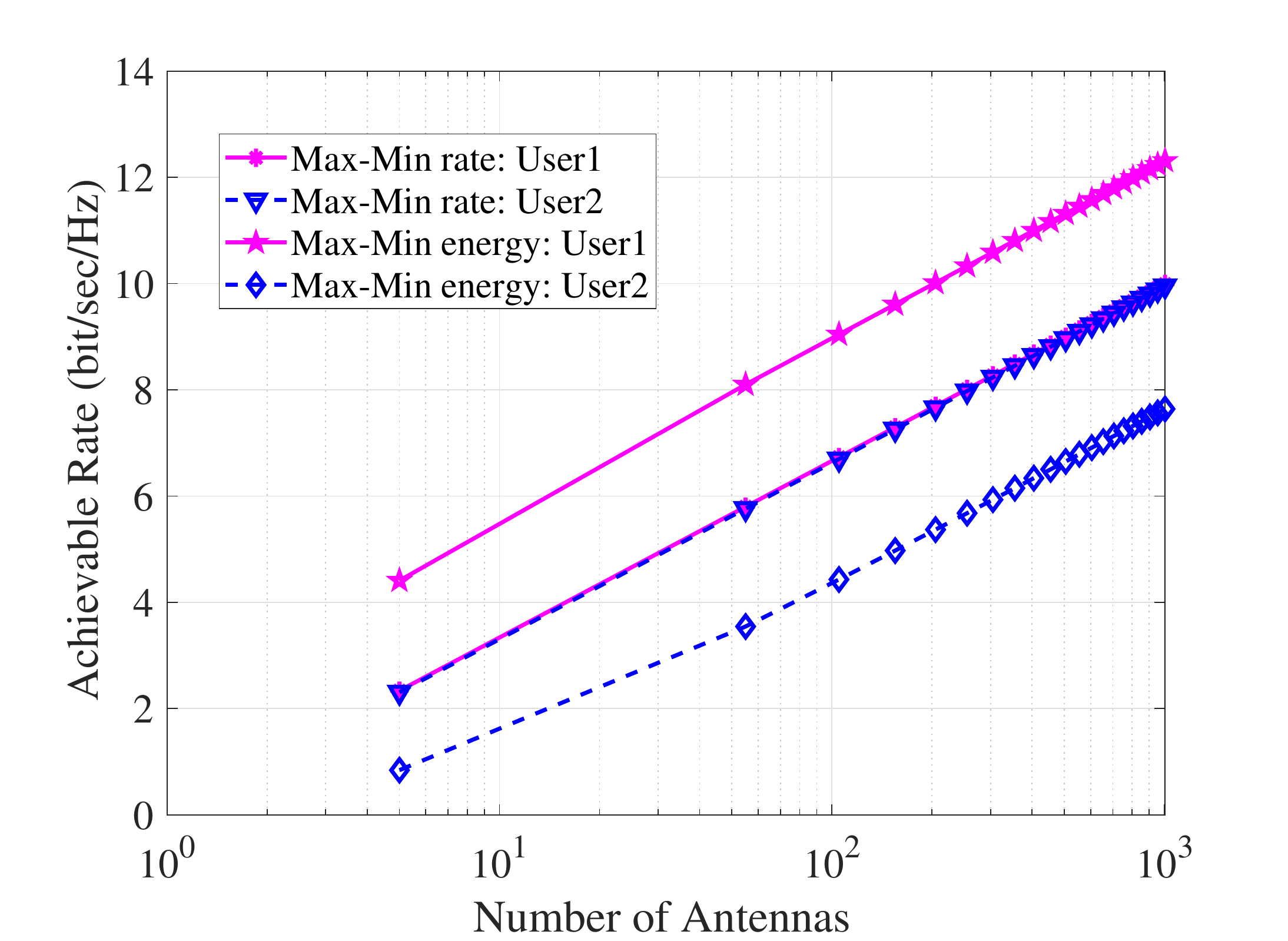}
	\caption{Achievable rate vs. $R$ for different optimization goals.}
	\label{fig:figure4}
\end{figure}

\begin{figure}
	\centering
	\includegraphics[width=0.85\linewidth]{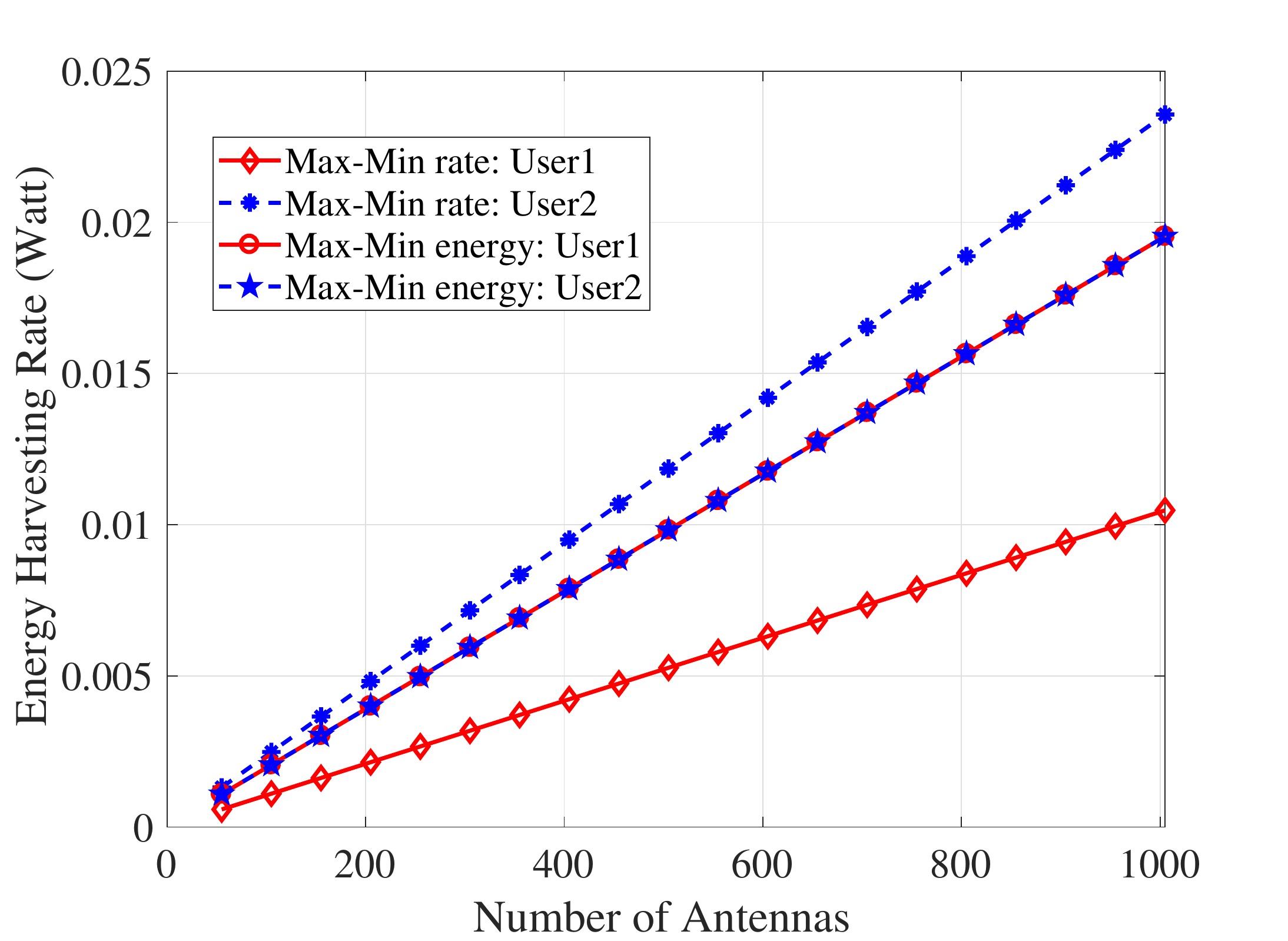}
	\caption{Energy harvesting rate vs. $M$ for different optimization goals.}
	\label{fig:figure5}
\end{figure}

Figs.~\ref{fig:figure4} and \ref{fig:figure5} compare our energy beamforming scheme, which maximizes the minimum rate for all tags subject to the power consumption constraint, and the existing energy beamforming design that aims to maximize the minimum energy harvesting rate. Fig.~\ref{fig:figure4} indicates that the minimum achievable rate of the proposed scheme is significantly higher than that of previous energy beamforming scheme. Besides, it shows that the proposed beamforming scheme can effectively guarantee the reliable communications and fairness of all tags. Even though the energy harvesting rate of tag $1$ is low, it can also reach the same achievable rate as tag $2$. The beamforming scheme which maximizes the minimum energy can ensure the energy harvesting rate of the two tags is equivalent. However, due to the double near-far effect \cite{yang2015throughput}, the rate of tag $2$ is $80\%$ lower than that of tag $1$, when $M=R=5$. Thus, it is wasteful for the portion of the energy that exceeds the circuit power consumption of the tag.

%\begin{figure}
%	\centering 
%	%并排几个图，就要写几个minipage
%	\begin{minipage}[b]{0.2 \textwidth} %所有minipage宽度之和要小于1，否则会自动变成竖排
%		\centering %图片局部居中
%		\includegraphics[width=1\textwidth]{Pictures/figure4} %此时的图片宽度比例是相对于这个minipage的，不是全局
%       	\caption{Achievable rate vs. $R$ for different optimization goals.}
%        \label{fig:figure4}
%	\end{minipage}
%	\begin{minipage}[b]{0.2 \textwidth} %所有minipage宽度之和要小于1，否则会自动变成竖排
%		\centering %图片局部居中
%		\includegraphics[width=1\textwidth]{Pictures/figure5}%此时的图片宽度比例是相对于这个minipage的，不是全局
%    	\caption{Energy harvesting rates vs. $M$ for different optimization goals.}
%        \label{fig:figure5}
%	\end{minipage}
%\end{figure}

\section{Related Work}
Significant progress has recently been made on the energy beamforming in wireless powered communication \cite{almradi2016information,chen2018beamforming,khan2018optimization,psomas2019energy}. To harness the benefits of energy beamforming, there have been many efforts made to enable energy beamforming in WPBC. Long et al. \cite{long2017transmit} focus on optimizing the transmit beamforming to maximize the sum rate of a cooperative WPBC system. Gong et al. \cite{gong2018backscatter} investigate energy beamforming in a relay WPBC system. Both these studies assume that F-CSI can be known to achieve energy beamforming. However, the closed-loop propagation and power-limited tags make it hard to obtain F-CSI. Instead, Yang et al. \cite{yang2015multi} performs energy beamforming via the estimated BS-CSI to improve WET efficiency. Departure from these studies, our work considers the beamforming design for both energy supply and data transfer to ensure communication performance. 

\section{Conclusion}

In this paper, we investigate the energy beamforming via estimated BS-CSI to improve the performance of multiuser WPBC system. To ensure uninterrupted communication, achievable rate and user fairness, we propose a beamforming scheme for power and information transfer in WPBC using the estimated BS-CSI. Additionally, we obtain the analytical expressions of the energy harvesting rate and lower bound on the ergodic achievable rate to optimize resource allocation for maximizing the minimum rate for all tags. Our results indicate that the proposed beamforming scheme offers tremendous performance gains, compared to state-of-the-art energy beamforming schemes. We hope that our investigation on the proposed energy beamforming scheme can provide some implications for future designs.

\begin{appendices} 
%\vspace{-0.1cm}
\section{Proof of Proposition 1} 
%\vspace{-0.1cm}
Based on \eqref{incident signal power} and $h_{kr}^{b} \sim \mathcal{CN}\left( 0, \beta_{k }\right)$, define the random variable $f=|h_{kr}^{b}|^2$, thus, the $f$ follows exponential distribution. According to the \cite{yang2015multi} (Lemma 3), we can obtain the expectation in \eqref{incident signal power} as follows.
%\vspace{-0.2cm}
\begin{equation}\label{expectation}
\vspace{-0.2cm}
\begin{aligned}
\phi\left(h_{kr}^{b}\right) = \frac{K\sigma^2}{\beta_{k}^2 \alpha p_{ce} \delta_{k}} \exp(\frac{K\sigma^2}{\beta_{k}^2 \alpha p_{ce} \delta_{k}}) \Gamma(0, \frac{K\sigma^2}{\beta_{k}^2 \alpha p_{ce} \delta_{k}}),
\end{aligned} 
\end{equation} 
where $	\Gamma(0,t)\triangleq\int_{t}^{\infty}u^{-1}\exp(-u)du$ and the bound of the Gamma function is \cite{abramowitz1964handbook}
%\vspace{-0.2cm}
\begin{equation}\label{Gamma bound}
%\vspace{-0.2cm}
\frac{1}{2} e^{-x} \ln\left( 1+ \frac{2}{x}\right) < \Gamma(0,x) = E_{1}(x) < e^{-x} \ln\left( 1+ \frac{1}{x}\right), \quad (x>0).
\end{equation}
Substituting \eqref{expectation} into \eqref{Gamma bound}, Proposition 1 is proved. 
%\vspace{-0.1cm}
\section{Proof of Proposition 2} 
%\vspace{-0.1cm}
We use the MRC detection, i.e. $\textbf{q}_{k} = \frac{\widehat{\textbf{h}}_{mk}}{\|\widehat{\textbf{h}}_{mk}\|}$, and $\widehat{\textbf{h}}_{i}^{b}= \frac{\widehat{\textbf{h}}_{mi}}{h_{mi}^{f}}$.
Define $\tilde{\psi}_{i} \triangleq \frac{\widehat{\textbf{h}}_{mk}^{H} \widehat{\textbf{h}}_{mi}}{\|\widehat{\textbf{h}}_{mk}\|}$, and $\tilde{\psi}_{i} \sim \mathcal{CN}\left(0,|h_{mi}^{f}|^2\left( \beta_{i}+\sigma_{\epsilon,mi}^2(h_{mi}^{f})\right) \right)$. Then the expectation in \eqref{lower boound} can be rewritten as 
%\vspace{-0.2cm}
\begin{equation}\label{MRC_EX}
\begin{aligned}
\mathbb{E} & \left\lbrace \frac{\sum_{i=1,i\neq k}^{K}  p_{i} \frac{1}{|h_{mi}^{f}|^2} |\tilde{\psi}_{i}|^2  + \sum_{i=1}^{K}  p_{i}  \sigma_{\epsilon,mi}^2(h_{mi}^{f})+ \tilde{\sigma}^2}  { p_{k} \frac{1}{|h_{mi}^{f}|^2} \|\widehat{\textbf{h}}_{mk}\|^2 }\right\rbrace\\
& = \bigg( \sum_{i=1,i\neq k}^{K}  p_{i} \mathbb{E}_{h_{mi}^{f}} \left\lbrace  \beta_{i}+\sigma_{\epsilon,mi}^2(h_{mi}^{f}) \right\rbrace\\
&\quad + \sum_{i=1}^{K}  p_{i} \mathbb{E}_{h_{mi}^{f}} \left\lbrace \sigma_{\epsilon,mi}^2(h_{mi}^{f}) \right\rbrace + \tilde{\sigma}^2 \bigg)
\mathbb{E} \left\lbrace  \frac{1}{ p_{k} \frac{1}{|h_{mi}^{f}|^2} \|\widehat{\textbf{h}}_{mk}\|^2} \right\rbrace\\
& =  \bigg( \sum_{i=1,i\neq k}^{K}  p_{i} \left(\beta_{i} + \frac{K \sigma^2}{\alpha p_{ce} \delta_{i}}\mathbb{E}_{h_{mi}^{f}} \left\lbrace\frac{1}{|h_{mi}^{f}|^2} \right\rbrace\right) \\
&\quad + \sum_{i=1}^{K}  p_{i} \frac{K \sigma^2}{\alpha p_{ce} \delta_{i}}\mathbb{E}_{h_{mi}^{f}} \left\lbrace\frac{1}{|h_{mi}^{f}|^2}\right\rbrace + \tilde{\sigma}^2 \bigg)
\mathbb{E} \left\lbrace \frac{1}{ p_{k} \frac{1}{|h_{mi}^{f}|^2} \|\widehat{\textbf{h}}_{mk}\|^2} \right\rbrace.\\
\end{aligned}
\end{equation}
calculate $\mathbb{E}_{h_{mi}^{f}} \left\lbrace \frac{1}{|h_{mi}^{f}|^2}\right\rbrace$, let $\tilde{x}_{i} = |h_{mi}^{f}|^2 \ge \tau (\tau>0)$, $\tilde{x}_{i} \sim \exp\left( \frac{1}{\beta_{i}}\right)$, then
%\vspace{-0.2cm}
\begin{equation}\label{ex}
\begin{aligned}
\mathbb{E}_{h_{mi}^{f}} \left\lbrace\frac{1}{|h_{mi}^{f}|^2}\right\rbrace  = \int_{\tau}^{\infty} \frac{1}{\beta_{i} \tilde{x}_{i}} e^{-\frac{\tilde{x}_{i}}{\beta_{i}}} d\tilde{x}_{i}= \frac{\Gamma(0,\frac{\tau}{\beta_{i}})}{\beta_{i}},
\end{aligned} 
\end{equation}
%\vspace{-0.1cm}
where $\Gamma(0,\frac{\tau}{\beta_{i}}) \triangleq \int_{\frac{\tau}{\beta_{i}}}^{\infty} t^{-1} e^{-t} dt$.

Since $\widehat{\textbf{h}}_{mk}|h_{mk}^{f} \sim \mathcal{CN}\left(\textbf{0}_{R},|h_{mk}^{f}|^2\left( \beta_{k}+\sigma_{\epsilon,mk}^2(h_{mk}^{f})\right) \textbf{I}_{R}\right) $, $Z_{mk} \triangleq \frac{2}{|h_{mk}^{f}|^2\left( \beta_{k}+\sigma_{\epsilon,mk}^2(h_{mk}^{f})\right)} \widehat{\textbf{h}}_{mk}^{H} \widehat{\textbf{h}}_{mk}$ follows central chi-square distribution with $2R$ degree of freedom. Thus the $\frac{1}{Z_{mk}} \sim$ Inv-$\chi^{2}(2R)$, and we have $\mathbb{E} \left\lbrace\frac{1}{Z_{mk}}\right\rbrace = \frac{1}{2(R - 1)}$. Then
%\vspace{-0.2cm}
\begin{equation}\label{ex2}
\begin{aligned}
\mathbb{E} &\left\lbrace \frac{1}{ p_{k} \frac{1}{|h_{mi}^{f}|^2} \|\widehat{\textbf{h}}_{mk}\|^2} \right\rbrace  = \mathbb{E} \left\lbrace \frac{2}{ p_{k} \left( \beta_{k} +\sigma_{\epsilon,mk}^2(h_{mk}^{f}) \right) Z_{mk}} \right\rbrace\\
& = \mathbb{E}_{h_{mk}^{f}} \left\lbrace \frac{1}{ p_{k} (R-1) \left( \beta_{k} +\sigma_{\epsilon,mk}^2(h_{mk}^{f}) \right) } \right\rbrace\\
& = \frac{1}{p_{k} (R-1)\beta_{k}} \mathbb{E}_{h_{mk}^{f}} \left\lbrace \frac{1}{1 + \frac{K\sigma^2}{|h_{mk}^{f}|^2 \alpha p_{ce} \delta_{k} \beta_{k}} } \right\rbrace\\
& = \frac{1}{p_{k} (R-1)\beta_{k}} \left( 1 - \mathbb{E}_{h_{mk}^{f}} \left\lbrace \frac{1}{\frac{|h_{mk}^{f}|^2\alpha p_{ce} \delta_{k}\beta_{k}}{K\sigma^2}+1} \right\rbrace\right).\\                     
\end{aligned} 
%\vspace{-0.2cm}
\end{equation}	
Then, we can utilize \eqref{expectation}, and substitute \eqref{ex} and \eqref{ex2} into \eqref{MRC_EX}. This completes the proof.

\end{appendices}

%\section*{Acknowledgement}
%The research was supported in part by the National Science Foundation of China under Grant , National Natural Science Foundation of China through grant , as well as the Fundamental Research Funds for the Central Universities with grand HUST:.

\bibliographystyle{IEEEtran}
\bibliography{IEEEabrv,./Energy_Beamforming}

\end{document}